\documentclass{ws-spin}
\usepackage{multicol,overcite}
\def\bib{B\kern-.05em{I}\kern-.025em{B}\kern-.08em}
\def\btex{B\kern-.05em{I}\kern-.025em{B}\kern-.08em\TeX}
\usepackage{amsfonts}
\usepackage{mathrsfs}
\usepackage{amsmath}
\usepackage{color}
\usepackage{graphicx}
\usepackage{bm}
\usepackage{amssymb}
\usepackage{epstopdf}
\usepackage{dcolumn}
\usepackage{longtable}
\usepackage[colorlinks=true, letterpaper=true, pdfstartview=FitV, linkcolor=blue, citecolor=blue, urlcolor=blue]{hyperref}
\begin{document}

\catchline{06}{1640003}{2016}{}{}

\markboth{Shengyuan A. Yang}{Dirac and Weyl Materials}

\title{Dirac and Weyl Materials: Fundamental Aspects and Some Spintronics Applications}

\author{Shengyuan A. Yang}

\address{Research Laboratory for Quantum Materials,
Singapore University of Technology and Design \\
Singapore 487372, Singapore\\
\email{shengyuan\_yang@sutd.edu.sg}}

\maketitle

\begin{history}
\end{history}

\begin{abstract}
Dirac and Weyl materials refer to a class of solid materials which host low-energy quasiparticle excitations that can be described by the Dirac and Weyl equations in relativistic quantum mechanics. Starting with the advent of graphene as the first prominent example, these materials have been attracting tremendous interest owing to their novel fundamental properties as well as the great potential for applications. Here we introduce the basic concepts and notions related to Dirac and Weyl materials and briefly review some recent works in this field, particularly on the conceptual development and the possible spintronics/pseudospintronics applications.
\end{abstract}


\begin{multicols}{2}
\section{Introduction}
In an attempt to construct a relativistic theory of electron, Dirac wrote down in 1928 the famous equation that is named after him.\cite{Dirac1928} In an explicitly covariant form, the Dirac equation looks like
\begin{equation}
i\hbar\gamma^\mu\partial_\mu\psi-mc\psi=0.
\end{equation}
A salient feature of this equation is that: the coefficients $\gamma^\mu$ are matrices required to satisfy the so-called Clifford algebra $\{\gamma^\mu,\gamma^\nu\}=g^{\mu\nu}$ with $g^{\mu\nu}$ the metric tensor, and hence the wave function $\psi$ must have a multi-component structure known as a spinor. The spinor structure, initially devised as a way to remedy the seemingly fatal flaw of an indefinite probability density associated with the Klein-Gordon equation, has profound consequences on fermionic properties.

One such consequence is the emergence of chirality. Consider the energy dispersion
\begin{equation}
E^2=c^2p^2+m^2c^4.
\end{equation}
In the ultra-relativistic regime, i.e. when the kinetic energy dominates over the mass $cp\gg mc^2$, the dispersion becomes linear. Meanwhile, the original Dirac equation can be decoupled into two pieces. With a proper choice of the $\gamma^\mu$ matrices (so-called chiral representation), the two decoupled equations can be written as
\begin{equation}\label{eq3}
i\hbar\partial_0 \psi_\text{R/L}= \pm c(\bm \sigma\cdot \hat{\bm p})\psi_\text{R/L},
\end{equation}
where $\hat{\bm p}=-i\hbar\bm \nabla$ is the momentum operator, $\bm\sigma$ is the vector of Pauli matrices, and $\psi_\text{R/L}$ here is a two-component spinor. The subscripts R and L indicate that there are two kinds of particles: the right- and left-handed particles, corresponding to the $\pm$ signs respectively, a property known as chirality. From Eq.(\ref{eq3}), one can observe that the chiral particle's spin is tied with its momentum. For positive energy states, a right-handed (left-handed) particle has its spin oriented along (opposite to) the momentum. Formally, one defines a helicity operator $\hat{h}= (\bm \sigma\cdot \hat{\bm p})/p$, such that the chirality eigenstates are also eigenstates of $\hat{h}$ with definite helicity. Note that this identification is only possible in the ultra-relativistic regime. At low energies, the right- and left-handed particles are coupled by the mass, hence chirality is not well-defined unless the particle is massless with $m=0$.
\\
\\
\begin{figurehere}
\centerline{
\includegraphics[width=9cm]{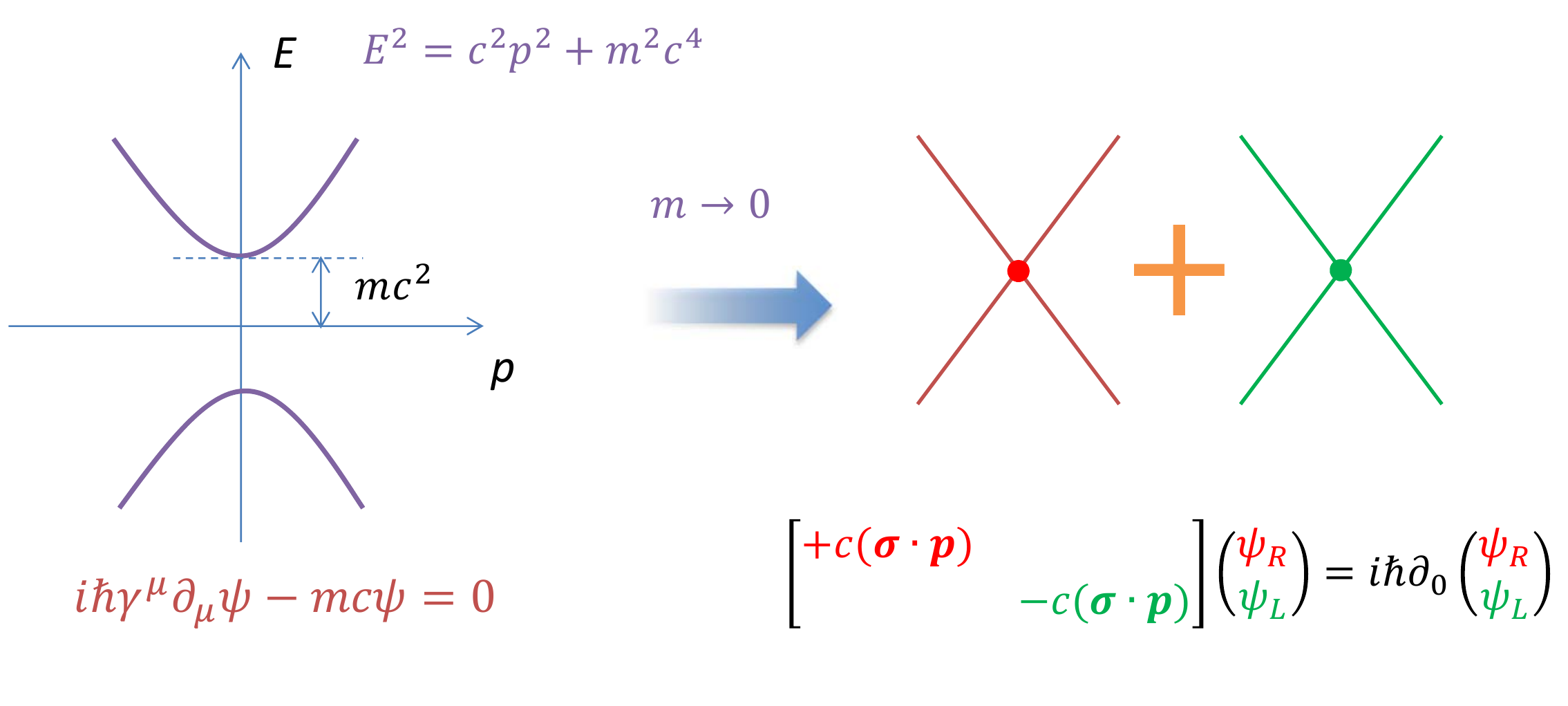}
}
\caption{Spectrum for a free Dirac particle (left) has a gap corresponding to the mass. For the massless case (right), the Dirac equation is decoupled into two Weyl fermions with definite and opposite chirality.}
\label{fig1}
\end{figurehere}

In other words, a massless particle is always in the ultra-relativistic regime. The Hamiltonian
\begin{equation}\label{Weyl}
{H}=\pm c(\bm \sigma\cdot \hat{\bm p})
\end{equation}
read from Eq.(\ref{eq3}) that describes a spin-$1/2$ massless fermion is known as the Weyl Hamiltonian, and the associated two-component spinor is known as the Weyl spinor. It features a well-defined chirality and a linear energy dispersion with two energy branches touching at a single point (at $p=0$), usually called the Weyl point.

Nevertheless, electrons do have a mass $\sim0.5$ MeV which is surely non-negligible on the typical energy scale of condensed matter physics. Hence at first glance, it may seem that the relativistic Dirac and Weyl models would have little to do with what happens for electrons in the solid state.
The point is that the solids present a complex environment where the electrons interact with the nuclei and with each other, and these interactions effectively modify the electronic properties. For many cases, strongly interacting electrons can be mapped to weakly interacting electron quasiparticles with a renormalized mass that can be very different from the bare mass. In crystalline solids, this corresponds to the formation of electronic band structure. One notes that the Weyl-like points would appear in the band structure if two non-degenerate bands cross each other at a single point. Then around that point, the quasiparticles are massless and described by the Weyl Hamiltonian with an emergent chirality.

Such linear band-crossing points are not uncommon in band structures. However, in condensed matter, the physics that we can probe is in a low-energy region around the Fermi level, therefore the crossing point has to be residing close to the Fermi level to manifest any effect, which poses a more stringent requirement. There has been studies in the past on these peculiar objects, particularly in the context of superfluid $^{3}$He.\cite{Volovik2009} A new surge of interest came about starting with the discovery of graphene in 2004.\cite{Novoselov2004}

\begin{figure*}
\begin{center}
\includegraphics[width=12cm]{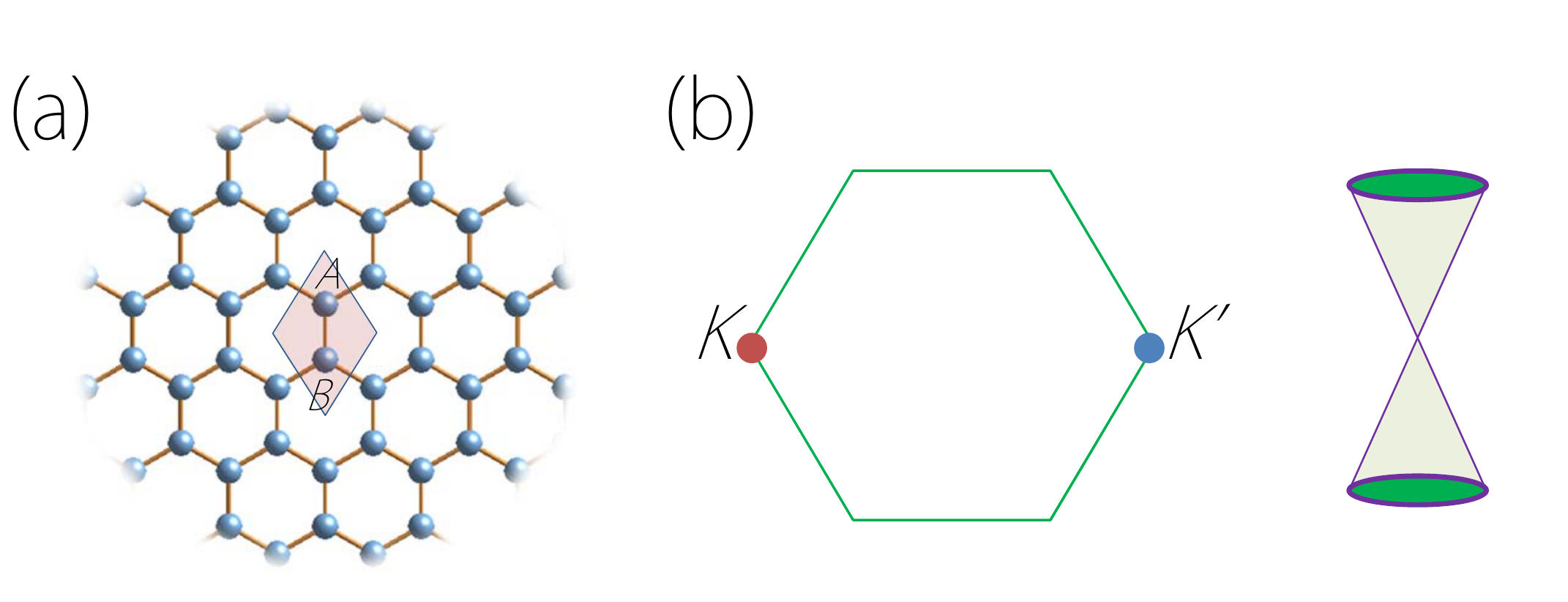}
\end{center}
\caption{(a) Graphene lattice with two basis atoms in a unit cell. The two sublattices are marked with A and B. (b) Two linear band-crossing points are located at the the corner points $K$ and $K'$ of the 2D Brillouin zone, with a conical dispersion shown schematically on the right.}
\label{fig2}
\end{figure*}

Graphene, a single layer of carbon atoms arranged in a honeycomb lattice structure, is the first and still the most important member of the fast-growing family of two-dimensional (2D) materials.\cite{Xu2013,Butler2013} Many remarkable properties of graphene can be attributed to its special band structure, in which the conduction and the valence bands linearly touch at two inequivalent points (denoted as $K$ and $K'$) at the hexagonal Brillouin zone corners (see Fig.~\ref{fig2}).\cite{CastroNeto2009} Without charge doping, the Fermi level would exactly cut through these points, making a point-like Fermi surface. Low-energy effective models can be obtained by $k\cdot p$ expansion around the two Fermi points, leading to Hamiltonians just like Eq.(\ref{Weyl}) in 2D. Therefore, each point represents a 2D Weyl point and the low-energy quasiparticle excitations are 2D chiral fermions. The low-energy band structure of graphene is quite clean with the linear dispersion extending over a wide energy range (over $1$ eV), making graphene an ideal platform for studying these relativistic chiral fermions.
\\
\\

Since graphene, a number of materials exhibiting Dirac/Weyl-type band structures have been identified. The concept is also generalized from 2D to 3D, leading to so-called Dirac and Weyl semimetals. These materials are expected to exhibit many intriguing effects, particularly the geometrical and topological effects connected with the chiral nature of the quasiparticles. The search for new Dirac/Weyl materials and the investigation on their properties are subjects at the frontier of the current condensed matter physics research.

In addition, as we mentioned, a Weyl particle's spin is tied with its momentum, representing a strongest type of spin-orbit coupling. For emergent Weyl fermions in condensed matter, one has to note that this ``spin" may or may not correspond to the real electron spin. For example, in the case of graphene, it is in fact a pseudospin of an orbital degree of freedom.\cite{CastroNeto2009} Furthermore, due to the fermion doubling theorem,\cite{Nielsen1981,Nielsen1981a} Weyl points must appear in pairs of opposite chirality in the band structure. For a pair of Fermi points separated in the reciprocal space, the freedom of an electron to reside at one of them --- the so-called valley degree of freedom ---
may also be treated as a pseudospin. All these suggest the great potential of Dirac/Weyl materials for applications in spintronics/pseudospintronics.

In this paper, we review some recent works in the research of Dirac/Weyl materials, which are primarily related to the aspects of conceptual development and proposals for spintronics/pseudospintronics applications, and the paper is organized into two parts corresponding to these two themes. In the short length, we must have missed out many interesting works and important aspects of this rapid-growing field. There have been several existing excellent reviews covering the field with emphasis on different topics.\cite{Pesin2012,Vafek2014,Wehling2014,Gibson2015,Yan2016} We hope that the current paper could help to bring certain new perspective towards a big picture.

\section{Conceptual Development}
In this section, we first introduce the basic concepts and notions related to Dirac/Weyl materials (Sec. 2.1). Then we review several examples of well-known 2D Dirac materials, focusing on their similarities and differences (Sec. 2.2). In Sec. 2.3, we discuss the possibility of new types of Dirac points in terms of their locations and dispersions. The realization of Dirac and Weyl physics in 3D leads to the concepts of 3D Dirac and Weyl semimetals. These materials and some of their generalizations will be discussed in Sec. 2.4.

\subsection{Basic concepts and notions}
In electronic band structures, Dirac/Weyl points correspond to special types of band crossing points. A Weyl point is where two bands cross with a linear dispersion. Hence, a Weyl point is two-fold degenerate. By analogy with the Dirac equation in the ultra-relativistic regime, a Dirac point can be regarded as consisting of two Weyl points of opposite chirality, i.e. with one right-handed Weyl point and one left-handed Weyl point sitting together, such that it is four-fold degenerate.

Generally, band-crossing points can be differentiated into two categories: the unprotected (accidental) crossing points and the ones protected by symmetry and/or topology.

For example, consider the standard textbook example of an electron in a 1D lattice.\cite{Ashcroft1976} In the so-called empty lattice approximation, i.e. when the lattice potential vanishes, there are many band crossings due to the band folding into the first Brillouin zone. These crossings are unprotected, because a perturbation such as a small lattice potential could remove them. Indeed, the Bragg reflections at periodic lattice potentials open gaps at these degeneracy points. This is how we learned the basic concepts of energy bands and bandgaps in a solid state physics course.

In comparison, consider the effective Hamiltonian describing a 3D Weyl point,
\begin{equation}\label{3DWeyl}
H(\bm k)=\pm v\bm k\cdot \bm \sigma,
\end{equation}
where $\bm k$ is the wave vector measured from the Weyl point, $v$ is like an effective speed of light, and the Pauli matrices represent the two degrees of freedom corresponding to the two crossing bands. One notes that any small perturbation cannot remove the crossing point: a perturbation expanded around the Weyl point takes the general form of $\delta_0 I_\mathrm{2\times 2}+\bm \delta\cdot\bm \sigma$ (the $\delta$'s may be expressed as Taylor expansions in the wave vector $\bm k$, and $I_\mathrm{2\times 2}$ is the $2\times 2$ identity matrix); the first term $ \delta_0 I_\mathrm{2\times 2}$ only shifts the Weyl point in energy, while the second term $\bm \delta\cdot\bm \sigma$ only shifts its location in $k$-space. Thus the 3D Weyl point is a protected band crossing. Its protection does not even require any extra symmetry (of course one needs the lattice translational symmetry to define a $k$-space), representing a kind of topologically stable object.

The situation changes if we squeeze the Weyl point to 2D. The loss of one dimension, e.g. removing the $k_z\sigma_z$ term in model (\ref{3DWeyl}), makes the band crossing vunerable against the perturbation of $\Delta\sigma_z$, which removes the 2D Weyl node (i.e. the crossing point) and opens a gap in the spectrum, just like the role of electron mass in the Dirac equation (hence such kind of terms are usually called mass terms). Thus a 2D Weyl node is not as robust as its 3D counterpart. Its stabilization would require the presence of certain symmetry. For the current case, a constraint of $\{H(\bm k), \sigma_z\}=0$ would forbid the $\Delta\sigma_z$ mass term. In the context of graphene, this constraint stands for the sublattice symmetry, i.e. the possibility to divide the graphene lattice into two sublattices that are equivalent (A and B as in Fig.~\ref{fig2}(a)). Hence, the 2D Weyl nodes in graphene are protected if the sublattice symmetry is preserved.

Before proceeding, we clarify the usage of certain terminologies. First, in literature, linearly dispersing band crossing points in 2D are almost invariably referred to as ``Dirac points". This probably stems from the usage in the early works on graphene. For graphene, each band crossing point at $K$ or $K'$ is described by a 2D Weyl Hamiltonian with a definite chirality, hence should be a Weyl point.
Note that the real electron spin here is a dummy degree of freedom due to the negligible spin-orbit coupling strength. Although it doubles the degeneracy of bands and of the crossing point, it does not affect the chirality. In fact, accounting for real spin, the crossing point may be regarded as two overlapping Weyl points with the same chirality.
The correspondence with Dirac is established only when we try to combine the quasiparticles at both $K$ and $K'$ points into a single model like
\begin{equation}
H=\left[
    \begin{array}{cc}
      H_K &  \\
       & H_{K'} \\
    \end{array}
  \right],
\end{equation}
which indeed contains a pair of Weyl points of opposite chirality but at different $k$-points. This is different from the later usage of this concept in 3D. Although it is conceptually more consistent to call the above objects Weyl points, in the following discussion, we choose to follow the common practice to use ``Dirac" for 2D (unless explicitly stated otherwise) and only differentiate ``Dirac" and ``Weyl" in 3D.

Second, as we have mentioned above, a 2D Dirac point can be gapped if no symmetry protects it. We shall see that
the gap opening for a 2D Dirac point may not be a bad thing, instead, it may generate even more interesting physics. In the 2D gapped case, we still speak of the ``Dirac point", which now refers to the band edge $k$-point, i.e. the $k$-point of the crossing assuming the gap is not open. This usage is due to the fact that although the Dirac band-crossing node is removed, there is still interesting Dirac-related physics from the gapped spectrum around the ``Dirac point".

Third, there are multiple types of spins. As discussed, the Pauli matrices $\sigma_i$ in the effective model (\ref{3DWeyl}) correspond to the two crossing bands; they are not necessarily connected to the real electron spin, and if this is the case, what they represent is typically called a pseudospin (or isospin). Nevertheless, in the effective Weyl Hamiltonian, they do act as the generators of the SO(3) group of rotations for spin-1/2, leading to a total angular momentum $\bm J$ for the quasiparticles around the Weyl node:\cite{Volovik2009}
\begin{equation}\label{AM}
J_i=L_i+\frac{\hbar}{2}\sigma_i,
\end{equation}
where $L_i$ is the orbital angular momentum. In this sense, the pseudospin $\sigma_i$ does play the role of an effective spin-1/2 for the low-energy Weyl quasiparticles. In addition, due to the fermion doubling theorem,\cite{Nielsen1981,Nielsen1981a} Weyl points must appear in pairs of opposite chirality. When these points are well-separated in $k$-space and degenerate in energy as required by certain symmetry (like $K$ and $K'$ points in graphene), we can treat the multiple low-energy subspaces around the points as another pseudospin degree of freedom. In semiconductor physics, the multiple energy-degenerate extremal points for conduction or valence band are referred to as ``valleys". By gapping a 2D or 3D Dirac node, we can have a special Dirac-type valley with many interesting properties. In analogy with conventional spintronics, it is possible to manipulate the valley pseudospin for information storage and processing, leading to the concept of valleytronics.

\begin{figure*}
\begin{center}
\includegraphics[width=12cm]{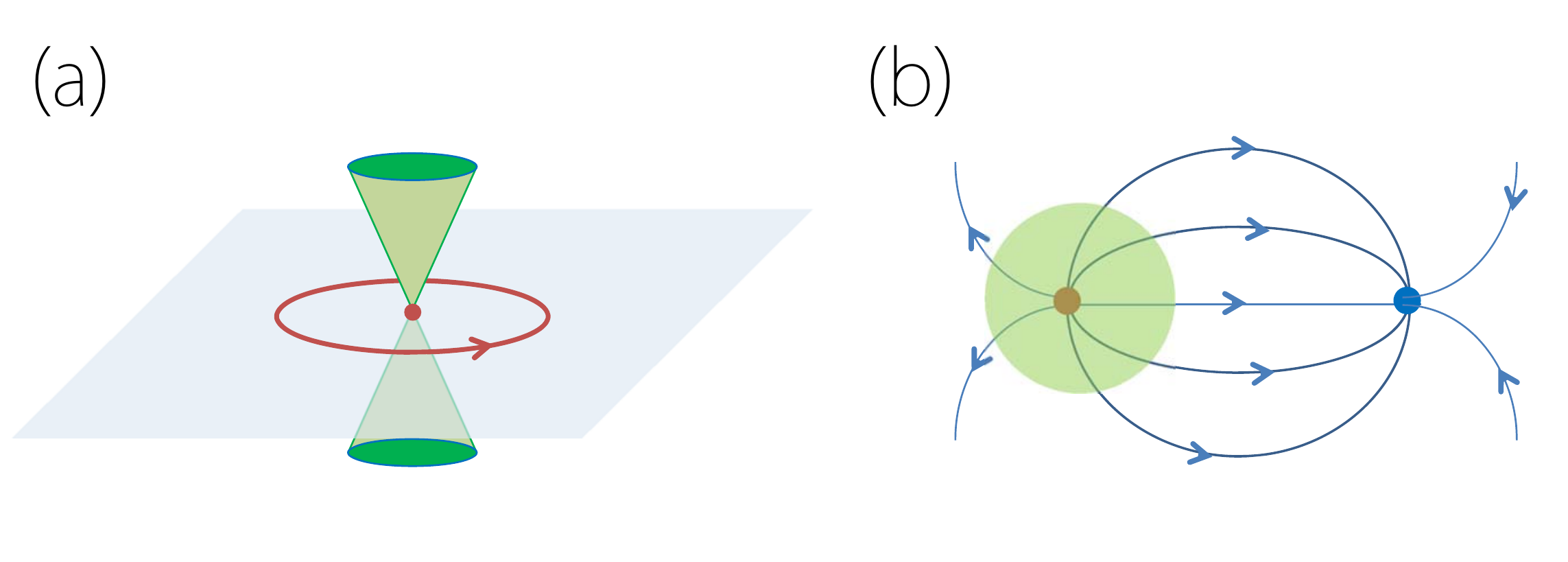}
\end{center}
\caption{(a) Integrating the Berry connection around a closed loop (marked in red color) around a 2D Dirac point gives a Berry phase of $\pm \pi$.
(b) 3D Weyl points (red and blue dots) act like source charges for the Berry curvature field. Integrating the Berry curvature on a closed surface (marked by the green color) gives the net topological charge enclosed.}
\label{fig3}
\end{figure*}

Back to the protected Dirac/Weyl point, mathematically, the protection is usually characterized by a topological invariant, an integer that labels configurations with different topologies. As the topology is for wave functions in $k$-space, one usually defines a $k$-space gauge potential, the so-called Berry connection,\cite{Xiao2010} (here we restrict the discussion to the Abelian case, i.e. no band degeneracy at the considered $k$-point)
\begin{equation}
\bm A_n(\bm k)=i\langle u_n|\nabla_{\bm k}u_n\rangle,
\end{equation}
which tracks the local phase variation of the energy eigenstates in $k$-space. Here $|u_n\rangle$ is the periodic part of the Bloch state and $n$ is the band index. Integration of $\bm A_n(\bm k)$ along a closed loop $C$ in $k$-space gives the well-known Berry phase:\cite{Berry1984}
\begin{equation}\label{BPhase}
\gamma_n=\oint_C \bm A_n(\bm k)\cdot d\bm \ell.
\end{equation}
Consider a 2D Dirac point,
\begin{equation}\label{2DDP}
H(\bm k)=v(\pm k_x\sigma_x+k_y\sigma_y),
\end{equation}
for a simple closed loop $C$ enclosing the band crossing point $\bm k=0$ (Fig.~\ref{fig3}(a)), the Berry phase $\gamma$ for the valence band is given by $\pm \pi$. In the presence of a chiral symmetry, the Berry phase must be quantized in units of $\pi$, thereby allowing the definition of a topological invariant: a 1D winding number $N_C=\gamma/\pi=\pm 1$, which corresponds to the chirality of the point.\cite{Schnyder2008}
Here chiral symmetry ($\mathcal{C}$) is a kind of unitary symmetry that anti-commutes with the Hamiltonian: $\{\mathcal{C}, H\}=0$. For model (\ref{2DDP}), $\mathcal{C}=\sigma_z$. One immediately observes that this is the same symmetry operator as the sublattice symmetry that we discussed above for graphene. Indeed, chiral symmetry often coincides with sublattice symmetry for crystalline solids.\cite{Schnyder2008} The requirement of a quantized $N_C$ protects the 2D Dirac node, because a gap opening will necessarily deviate $N_C$ from an integer value.

There is also a gauge field associated with the gauge potential, known as the Berry curvature,\cite{Xiao2010}
\begin{equation}
\bm \Omega_n(\bm k)=\nabla_{\bm k}\times \bm A_n,
\end{equation}
which is like a magnetic field for the vector potential $\bm A_n$ in $k$-space. For a 3D Weyl point as in model (\ref{3DWeyl}), direct calculation shows that $\bm \Omega(\bm k)=\pm \bm k/(2k^3)$ for the valence band, just like fields generated by a monopole charge. Hence Weyl points can be viewed as source or drain (depending on their chirality) for the Berry curvature field lines. We can define a topological charge (invariant) for a 3D Weyl point via a similar Gauss law
\begin{equation}\label{Chern}
N_S=\frac{1}{2\pi}\oint_S \bm \Omega(\bm k)\cdot d\bm \sigma,
\end{equation}
where $S$ is a simple closed surface enclosing the Weyl point (Fig.~\ref{fig3}(b)). The $N_S$ defined on a 2D closed surface is also known as the Chern number. One can see that $N_S=\pm 1$ for a Weyl point, corresponding to its chirality.

\subsection{2D Dirac materials}
\begin{figure*}
\begin{center}
\includegraphics[width=12cm]{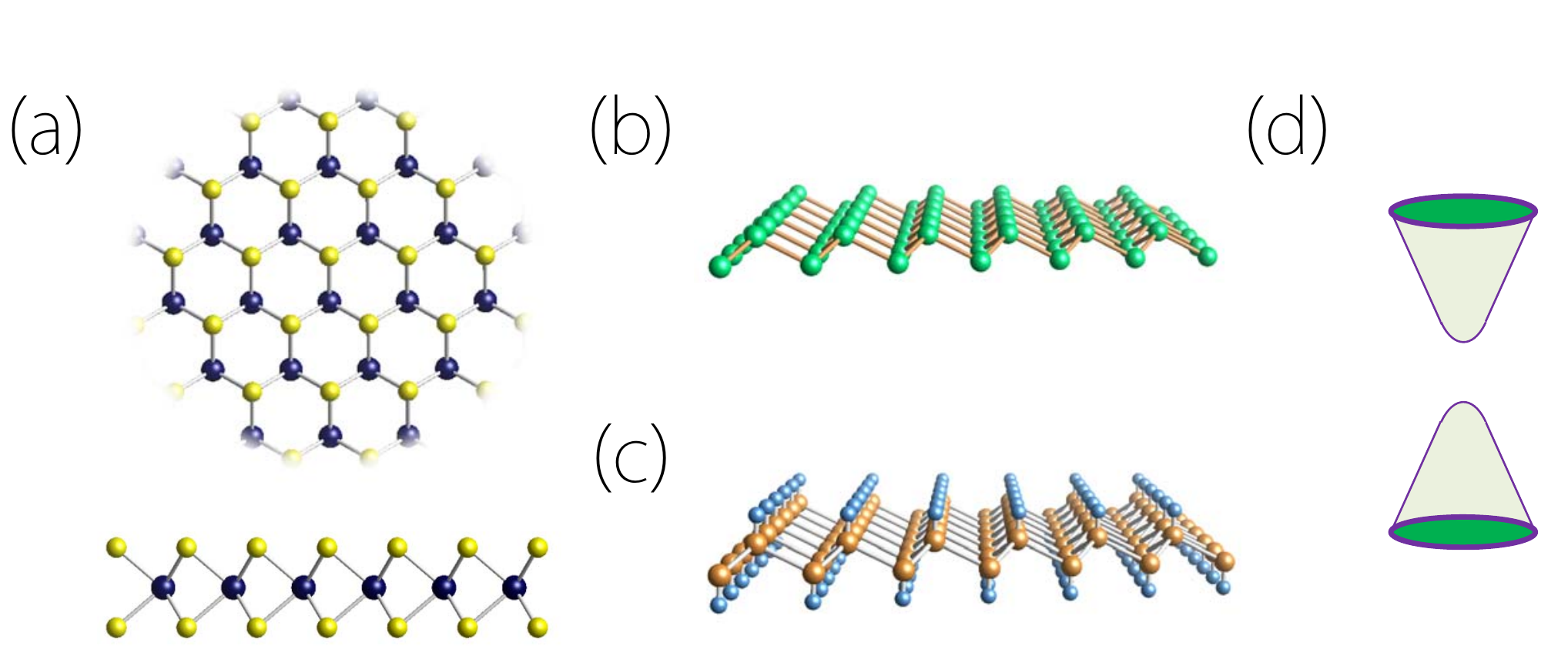}
\end{center}
\caption{Lattice structures of (a) transition metal dichalcogendies (top view and side view), (b) buckled honeycomb lattice materails (such as silicene, germanene etc.), and (c) functionalized 2D honeycomb lattice materials (such as BiH and SnI etc.). (d) Gap opened at a 2D Dirac point due to various symmetry breaking effects. Figs.(a-c) illustrated by Shan Guan.}
\label{fig4}
\end{figure*}

The first and most prominent example of 2D Dirac materials is graphene, a single layer of carbon atoms with honeycomb lattice structure. There, the two linearly touching bands are mainly from carbon $p_z$ orbitals (so-called $\pi$ bands). The Dirac points are located at high-symmetry points $K$ and $K'$, the two corner points of a hexagonal Brillouin zone. The effective Hamiltonian expanded around the Dirac points can be written as\cite{CastroNeto2009}
\begin{equation}\label{graphene}
H(\bm k)=v(\tau_z k_x\sigma_x+k_y\sigma_y),
\end{equation}
where $\tau_z=\pm$ labels the two valleys and also indicates their chirality. The two basis for the pseudospin $\sigma_i$ are $p_z$ orbitals located respectively on A and B sublattices. As we mentioned in the previous section, the sublattice symmetry provides a protection of the Dirac node from gap opening. Different ways to break the sublattice symmetry can give rise to different mass terms and drive the system to different phases.

For example, consider the following two mass terms that preserve the time reversal symmetry: $\sigma_z$ and $\tau_z \sigma_z s_z$. The first term beaks the sublattice symmetry as well as the inversion symmetry (represented by $\mathcal{P}=\sigma_x$); it drives the system to a so-called quantum valley Hall (QVH) phase. The second term beaks the sublattice symmetry but preserves the inversion symmetry; it couples the real spin $s_z$ to the orbital pseudospins and drives the system to the quantum spin Hall (QSH) phase.\cite{Kane2005,Kane2005a}

The $\sigma_z$ mass term can be induced in graphene by interaction with certain substrate, or more directly, by making the two sublattices occupied by different atoms, as in the family of 2D transition metal dichalcogenides (TMDs).\cite{Wang2012} When viewed from top, the 1H phase of 2D monolayer TMDs MX$_2$ (M=Mo, W; X=S, Se) has a similar honeycomb lattice structure but with different atoms M and X on the two sublattices (Fig.~\ref{fig4}(a)). This strong inversion symmetry breaking leads to a large mass term on the order of 1 eV. Besides, there is sizable spin-orbit coupling (SOC) effect (from the transition metal $d$-orbitals) for the valence band top, making the low-energy model a spin-split massive Dirac type\cite{Xiao2012}
\begin{equation}
H(\bm k)=v(\tau_z k_x\sigma_x+k_y\sigma_y)+\Delta\sigma_z+\lambda\tau_z s_z(1-\sigma_z),
\end{equation}
where $\lambda$ is the effective SOC strength on the order of $0.01\sim0.1$ eV depending on the TMD material.

The spin-orbit-coupled $\tau_z \sigma_z s_z$ term is allowed by symmetry in graphene. It is usually called the intrinsic SOC term.\cite{Kane2005} It drives the system to an interesting QSH insulator phase characterized by a topological $\mathbb{Z}_2$ invariant ($\mathbb{Z}_2=1$ for the QSH phase).\cite{Hasan2010,Qi2011} It is insulating in the bulk but conducting on the edge with a pair of gapless spin-helical edge channels. The two edge bands cross linearly with each other, and ``spin-helical" means the spin of the edge state is tied with its momentum. So the edge may also be regarded as a 1D Dirac system.

However, the intrinsic SOC term in graphene is negligibly small, due to the very weak atomic SOC strength $\xi$ for carbon and also due to the planar structure which makes the coupling as from a virtual process second order in $\xi/V$ ($V$ is a large energy scale on the order of the Dirac band width).\cite{Yao2007,Min2006} The $\tau_z \sigma_z s_z$ term can be enhanced by substituting carbon with heavier elements in the same group, leading to 2D silicene, germanene, stanene, and Pb layers.\cite{Liu2011b,Xu2013a,Lu2016} Like graphene, the low-energy bands at $K$/$K'$ points of these materials are of Dirac type and mainly from the atomic $p_z$ orbitals. The enhancement comes from two factors:\cite{Liu2011} (1) the atomic SOC strength increases as the fourth power of the atomic number; (2) the lattice structure becomes buckled (with A and B sites on different atomic planes) (Fig.~\ref{fig4}(b)), making it possible to couple $\pi$ and $\sigma$ states directly, an ingredient needed to make the effective SOC from a virtual process first order in $\xi/V$. Meanwhile, the buckled structure also allows an easy control of the $\sigma_z$ mass term via a perpendicular $E$-field, leading to interesting competition effects.\cite{Ezawa2012c,Tsai2013,Yang2015a}

Covalent functionalization of the 2D layers with e.g. hydrogen or halide groups can further increase the global QSH insulator gap. However, this saturates the $p_z$ orbitals. As a result, the states at $K$/$K'$ points are pushed to higher energies, completely removing the Dirac bands.\cite{Xu2013a,Lu2016} It is noted that for similarly functionalized 2D honeycomb lattices of group-Va elements Sb and Bi (Fig.~\ref{fig4}(c)), the Dirac bands at $K$/$K'$ can be preserved.\cite{Song2014,Liu2014g} Furthermore, the low-energy states are from atomic $p_x$/$p_y$ orbitals, which permits an on-site type effective SOC, leading to a huge QSH gap on the order of 1 eV.\cite{Liu2014g}

In addition, 2D Dirac fermions also appear at the surface of 3D topological insulators.\cite{Hasan2010,Qi2011,Shen2013} Such system is quite different from the above examples in the following aspects. For each surface, there are an odd number of Dirac points (with their partners on the opposite surface), which are pinned at the time reversal invariant $k$-points (frequently the $\Gamma$ point) of the surface Brillouin zone, and its energy can vary depending on the surface condition. For some materials, the node can even be buried into the bulk bands. In addtion, the Pauli matrices in the Weyl Hamiltonian here stand for the real spin.

\subsection{New types of 2D Dirac points}

One can see that in the above-mentioned materials, Dirac points all appear at high-symmetry points (like $K$ and $K'$). As long as the crystalline symmetry is preserved, they are pinned at these points. Further, due to the three-fold rotational symmetry at $K$ and $K'$ points, the dispersion must be isotropic to the leading order in $k$.

Is there any 2D material that can host Dirac points not located at high-symmetry points? Examples have been discovered in a few nanostructured materials based on first-principles calculations,\cite{Wang2015} including graphyne,\cite{Malko2012} some carbon and boron allotropes with rectangular structures,\cite{Xu2014,Zhou2014} and germanene on Al(111) structure.\cite{Liu2015a} In these materials, there are Dirac points appearing on high-symmetry lines. With preserved crystalline symmetry, these Dirac points are free to move along the line. And because of the reduced symmetry, their energy dispersions become anisotropic. For example, the germanene on Al(111) structure has totally eight Dirac points, of which six are located on high symmetry lines $M$-$K$($K'$).\cite{Liu2015a} The dispersion at the six points are characterized by two very different Fermi velocities for the two directions along and perpendicular to the line.

Nonetheless, for all the above cases, there is still some remaining symmetry constraining the location and dispersion of the Dirac point. Is it possible to have a truly unpinned 2D Dirac point, located at a generic $k$-point?

\begin{figure*}
\begin{center}
\includegraphics[width=12cm]{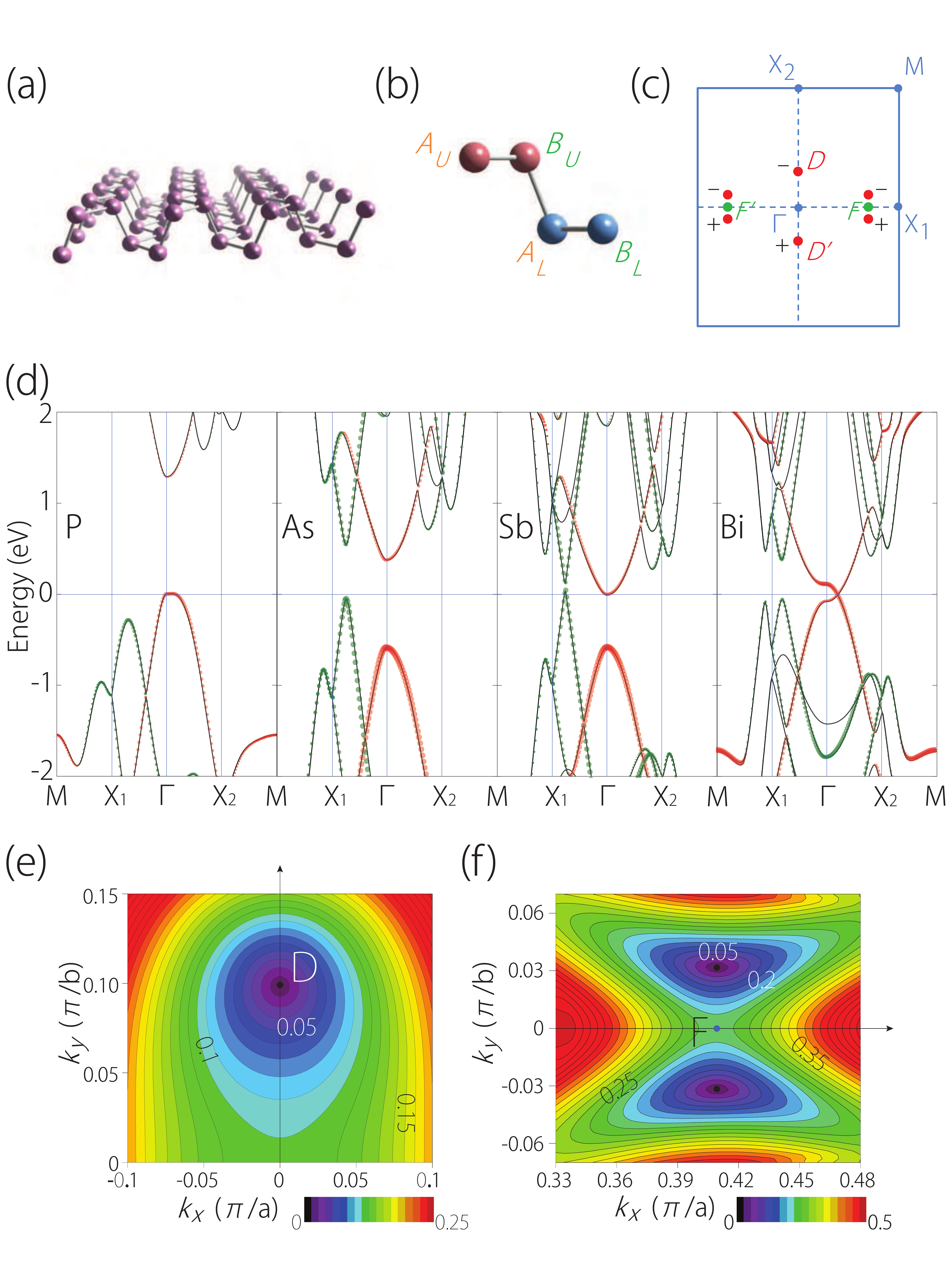}
\end{center}
\caption[]
{(a) Phosphorene lattice and (b) four-atom basis in a unit cell. (c) Red dots stand for the Dirac points: one pair is on high-symmetry lines and the other two pairs are fully-unpinned at generic $k$-points. (d) Band structures for group-Va elements with the phosphorene lattice structure. (e) and (f) are the energy dispersions around points $D$ and $F$ in (c), respectively.
 Figure adapted with permission from Ref.~\refcite{Lu2016a}.}
\label{fig5}
\end{figure*}

The answer is yes. The first example is discovered in the 2D phosphorene lattice.\cite{Lu2016a} Phosphorene is a single layer of the black phosphorus crystal.\cite{Li2014} In a side view, it has an armchair-like puckered lattice structure. Its rectangular unit cell has four atomic sites which can be labeled as in Fig.~\ref{fig5}(b). The phosphorene band structure shows a typical semiconductor with a sizable bandgap $>1$ eV at the $\Gamma$ point, without any trace of a Dirac point (Fig.~\ref{fig5}(d)). Interesting observations have been made by replacing P atoms with heavier elements (As, Sb, and Bi) from the same group, for which the direct gap at the $\Gamma$ point decreases and eventually a pair of Dirac points appear for Bi near the $\Gamma$ point along the $\Gamma$-$X_2$ line. More importantly, another two pairs of Dirac points emerge at generic $k$-points near the $\Gamma$-$X_1$ line (see Fig.~\ref{fig5}(c)) These Dirac points are fully unpinned: they can move around in the Brillouin zone without any lattice symmetry breaking. And due to the reduced symmetry, their dispersions become highly anisotropic (Fig.~\ref{fig5}(e,f)).

Another interesting aspect of this system comes from its four-atom basis. In the absence of SOC, each of the six Dirac crossings is protected by the $\pi$-quantized Berry phase (or 1D winding number) as in Eq.(\ref{BPhase}) due to the \{A\}-\{B\} sublattice symmetry. One observes that this symmetry can in turn be enforced by a multiple of point group symmetries including $i$ (inversion), $c_{2y}$, and $c_{2z}$ (180$^\circ$ rotation around $\hat y$ and $\hat z$, respectively). It has been shown that each of the three symmetries could protect the Dirac nodes independent of the other two.\cite{Lu2016a} For example, a vertical electric field breaks $i$ and $c_{2y}$ but not $c_{2z}$, therefore all the Dirac nodes survive, only their locations are shifted.

Without SOC, the Dirac points can only generate or annihilate in pairs of opposite chirality, a process corresponding to a quantum/topological phase transition. SOC generally breaks sublattice symmetry. For heavy elements such as Sb and Bi, the SOC effect is strong, leading to gap opening at the Dirac nodes. For the Bi lattice, this drives the system into a QSH insulator phase.\cite{Lu2014} On the other hand, for light elements such as P, it was shown that the fully unpinned Dirac points can still be realized by uniaxial strains,\cite{Lu2016a} implying that such Dirac points are a result of the specific lattice structure.

In most cases, SOC gaps 2D Dirac nodes. Recently, another new type of 2D Dirac points are discovered, which are robust against SOC. In a theoretical study,\cite{Young2015} Young and Kane showed that nonsymmorphic space group symmetries could protect Dirac nodes at certain high-symmetry $k$-points, and may lead to an interesting hourglass type dispersion (which was also found recently at the surface of a class of topological insulator materials.\cite{Wang2016a}) Similar Dirac nodes were also discovered in certain charge/spin density wave states, which are protected by nonsymmorphic symmetries against a dynamically generated SOC, as shown by Venderbos.\cite{Venderbos2016} These special states are at the boundary of a topological phase transition, making them suitable platforms for engineering various topological phases via symmetry breaking.

\subsection{3D Dirac/Weyl semimetals and beyond}

In 3D, the realization of a Weyl point requires the breaking of either time reversal symmetry ($\mathcal{T}$) or inversion symmetry ($\mathcal{P}$). To see this, first consider how these symmetry operations affect the Berry curvature and the Weyl point chirality. $\mathcal{T}$ maps $\bm k$ to $-\bm k$, and at the same time reverses the $k$-space ``magnetic field" $\bm \Omega$, i.e. $\bm \Omega(-\bm k)=-\bm \Omega(\bm k)$. Hence, a Weyl point $\bm k$ under $\mathcal{T}$ is mapped to another Weyl point at $-\bm k$ with the same chirality. In contrast, under $\mathcal{P}$, $\bm \Omega(-\bm k)=\bm \Omega(\bm k)$, and a Weyl point $\bm k$ will be mapped to another Weyl point at $-\bm k$ with reversed chirality. As a result, if the system enjoys both symmetries, the net Berry curvature field must vanish, so that there is no isolated Weyl points (which are source charges for $\bm \Omega$). Alternatively, one can argue that for a possible Weyl point, the combined operation $\mathcal{PT}$ will lead to a second Weyl point with opposite chirality at the same $k$-point, making it a Dirac point instead of Weyl.

Symmetry also puts constraint on the number of Weyl points. With preserved $\mathcal{T}$, a Weyl point and its time reversal partner have the same chirality, there must be another two Weyl points with reversed chirality to fulfill the fermion doubling theorem. Therefore the total number in this case must be a multiple of four. Only when $\mathcal{T}$ is broken can we have two Weyl points, the smallest number required by the fermion doubling theorem.

The possibility to have an extended phase with a Weyl point Fermi surface was pointed out by Murakami in studying the phase transition from a topological insulator to a trivial insulator, with $\mathcal{P}$ symmetry broken.\cite{Murakami2007} The term ``Weyl semimetal" was proposed by Wan \emph{et al.} with the first solid state example in a family of pyrochlore iridates material.\cite{Wan2011} They showed by first-principles calculations that the material Y$_2$Ir$_2$O$_7$ in the semimetal phase has 24 Weyl points at Fermi level. More importantly, they pointed out the surface correspondence of the nontrivial topology in the bulk: there exist topological Fermi arcs on the surface of the system, connecting pairs of surface-projected Weyl points with opposite chirality.

Since then, Weyl points have been proposed in a variety of systems.\cite{Burkov2011,Lu2013,Gong2011} In 2015, proposals of Weyl semimetal phase in a family of transition metal monophosphides were confirmed in experiment.\cite{Weng2015,Huang2015,Xu2015a,Lv2015,Yang2015c,Lv2015a} The first-principles calculations showed that around the Fermi level, there are 24 Weyl points in the bulk with fairly complicated surface Fermi arc patterns. Through angle-resolved photoemission spectroscopy (ARPES), the bulk Weyl nodes as well as the surface Fermi arcs can be imaged, which compare well with the first-principles calculation result.

The concept of ``Dirac semimetal" was first proposed by Young \emph{et al.}.\cite{Young2012} They pointed out the need of space group symmetries to stablize the Dirac point, a point consisting of two Weyl nodes of opposite chirality. An example is predicted in a metastable phase of $\beta$-cristobalite BiO$_2$. The first two experimentally confirmed Dirac semimetals are based on the proposals by Wang \emph{et al.}, in Na$_3$Bi and Cd$_3$As$_2$ materials.\cite{Wang2012b,Wang2013b} In these materials, there are two Dirac points located on the $k_z$-axis near the $\Gamma$ point, protected by the rotational symmetry of the respective lattice.

Despite their different structures and compositions, the low-energy behavior of Na$_3$Bi and Cd$_3$As$_2$ can be well captured using the following effective model expanded at the $\Gamma$ point,\cite{Wang2012b,Wang2013b}
\begin{equation}\label{DSMH}
H(\bm k)=\varepsilon_0(\bm k)I_{4\times 4}+\left[
                                \begin{array}{cccc}
                                  M(\bm k) & Ak_+ &  &  \\
                                  Ak_- & -M(\bm k) &  &  \\
                                   &  & M(\bm k) & -Ak_- \\
                                   &  & -Ak_+ & -M(\bm k) \\
                                \end{array}
                              \right]
\end{equation}
with $M(\bm k)=M_0-M_1k_z^2-M_2(k_x^2+k_y^2)$, $k_\pm=k_x\pm ik_y$, $A$ and $M_i$'s are material parameters, and certain higher order and symmetry-breaking terms are neglected.  With $M_0,M_1,M_2<0$, the two Dirac points are located at $k_z=\pm k_D$ with $k_D=\sqrt{M_0/M_1}$.

One notes that for each fixed $k_z$, the model (\ref{DSMH}) has the same form as that for the 2D QSH insulator in HgCdTe quantum wells.\cite{Bernevig2006,Shen2013} This analogy can provide us with a simple picture to understand the nontrivial topology and surface Fermi arcs of these Dirac semimetals. If we take a 2D slice of the Brillouin zone perpendicular to $k_z$-axis, the 2D $\mathbb{Z}_2$ invariant of the slice changes from 0 to 1 and back to 0 as it crosses the two Dirac points successively (see Fig.~\ref{fig6}). Each $\mathbb{Z}_2=1$ slice contributes a pair of topological boundary states, hence forming two surface Fermi arcs connecting the projected Dirac points on the side surfaces. These surface Fermi arcs have been found in first-principles calculations and been observed in ARPES experiment.\cite{Xu2015}
\\
\\
\begin{figurehere}
\centerline{
\includegraphics[width=8cm]{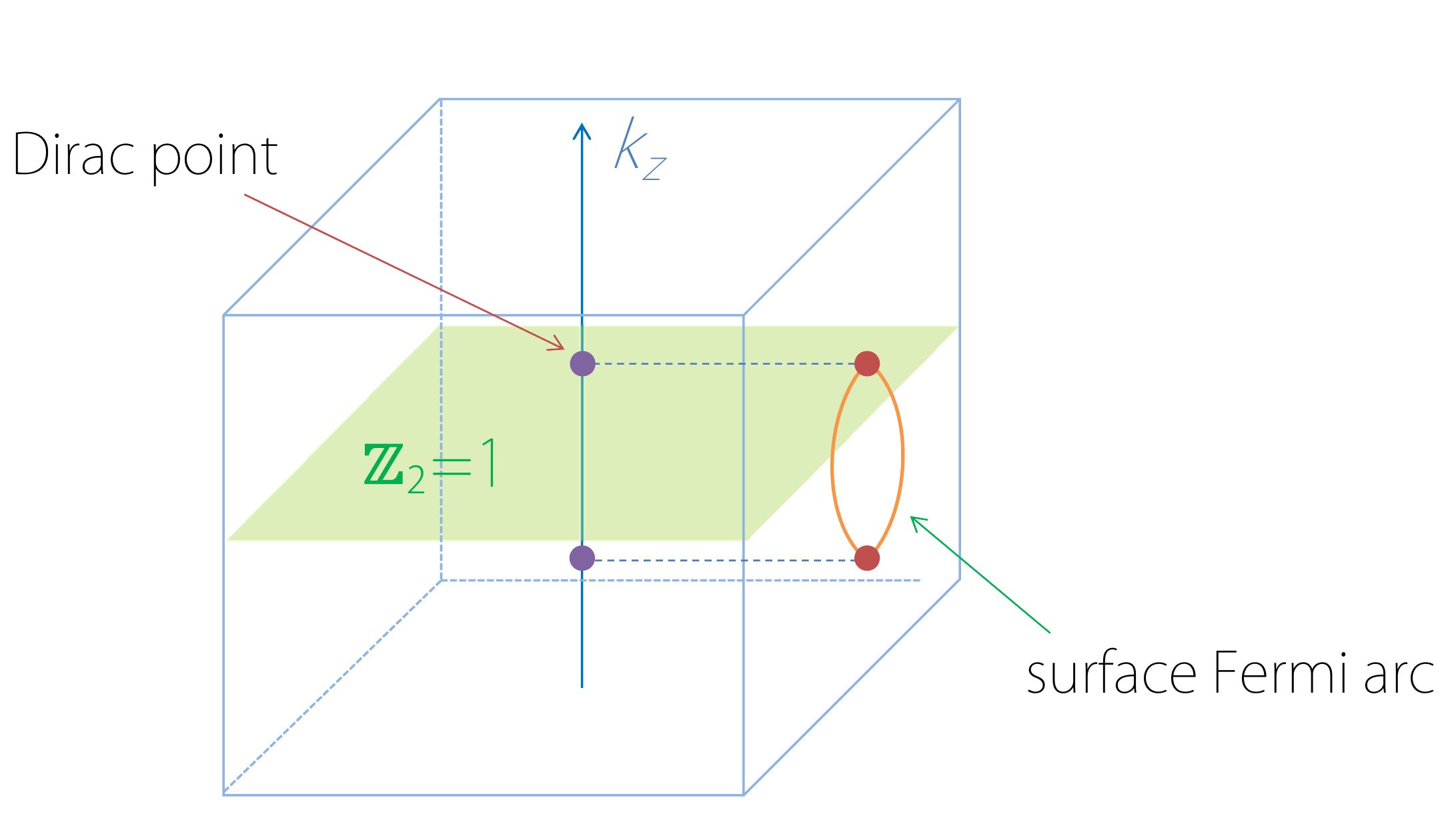}
}
\caption[]{Schematic figure for the Dirac semimetal phase in Na$_3$Bi and Cd$_3$As$_2$. The bandgap is inverted in between the two bulk Dirac points. A slice with fixed $k_z$ between the Dirac points can be viewed as an effective 2D system with a nontrivial $\mathbb{Z}_2$ invariant, dictating the presence of surface Fermi arcs. Figure adapted with permission from Ref.~\refcite{Xiao2015}.}
\label{fig6}
\end{figurehere}

There are many interesting possibilities to go beyond the Dirac/Weyl semimetals. The dispersion around a band crossing point may be quadratic instead of linear along certain directions, giving a Chern number of 2, leading to so-called double Weyl point.\cite{Xu2011a,Fang2012} The dispersion may be strongly tilted to make the Fermi surface topology transfer from a point to a line or a surface, leading to so-called type-II Weyl point.\cite{Soluyanov2015,Xu2015b} The dimensionality of the band crossing manifold may also vary from a point (0D) to a loop (1D) and to a surface (2D).
\\
\\
\begin{figurehere}
\centerline{
\includegraphics[width=8cm]{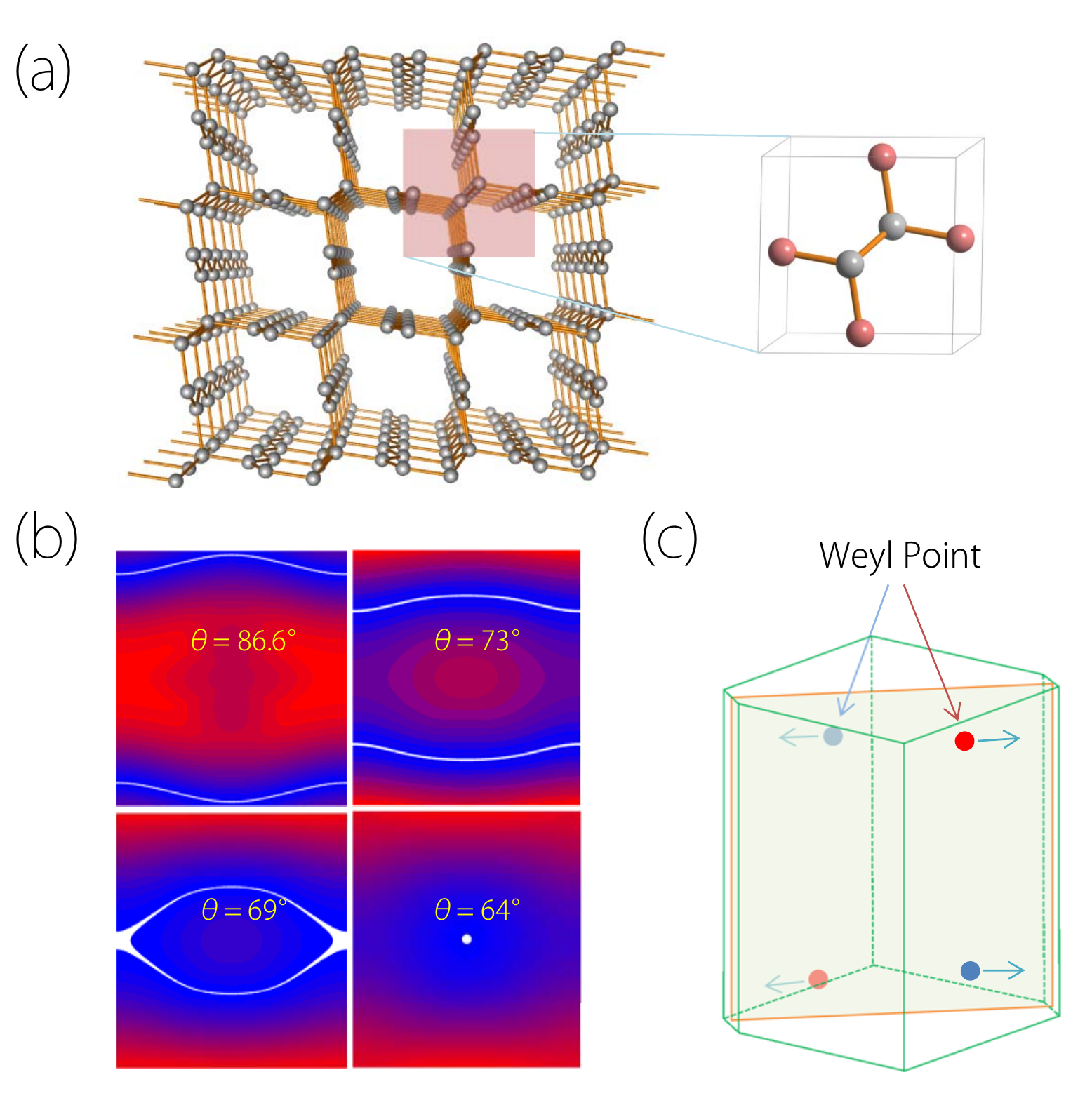}
}
\caption[]{(a) Lattice structure of a 3D carbon allotrope with six atoms in a unit cell. (b) Two Weyl loops appear in the (110) mirror invariant plane of the Brillouin zone. Under strain, the two loops shift, merge into a single loop, and finally annihilate. (c) Under inversion symmetry breaking, each loop transforms into a pair of Weyl points. Figure adapted with permission from Ref.~\refcite{Chen2015}. Copyright 2015 American Chemical Society.}
\label{fig7}
\end{figurehere}

The Dirac/Weyl loops have been proposed in quite a few solid materials.\cite{Phillips2014,Weng2015a,Mullen2015,Chen2015,Yu2015,Kim2015a,Chen2015a,Liang2016} Like Dirac points, their stability requires the protection by certain crystalline symmetry.\cite{Zhao2013c,Fang2015} Notably, there has been several proposal of nodal loops in carbon allotropes.\cite{Weng2015a,Mullen2015,Chen2015} Due to their freedom to form various hybridized bonds, carbon atoms can be assembled into a vast number of allotropic forms. And due to the negligible SOC strength of carbon, possible sublattice symmetry can be preserved more easily without worrying about the detrimental effect from SOC. Note that because the real spin is a dummy degree of freedom introducing a trivial degeneracy factor of two, in line with the discussion in Sec. 2.1, it is more appropriate to call the following band crossing as Weyl rather than Dirac.
An example is shown in Fig.~\ref{fig7}, which can be viewed as an interpenetrated graphene network, with a simple unit cell consisting of six carbon atoms.\cite{Chen2015} A pair of Weyl loops appear around Fermi level, residing in a vertical mirror plane. The loops are distinct in the following sense: each loop traverses the whole Brillouin zone, so that it cannot continuously shrink in size; with preserved sublattice symmetry, each loop is protected by a 1D winding number (as in Eq.(\ref{BPhase}) for another loop encircling this loop), and can only annihilate by merging with its partner. The latter scenario is observed when applying strain to the system. During the process, the two loops first merge into a single one, which then shrinks into a point and finally becomes fully gapped (Fig.~\ref{fig7}(b)). By breaking the inversion symmetry (and the sublattice symmetry), the Weyl loops can transform into Weyl points (Fig.~\ref{fig7}(c)).

It is also possible to have a Weyl surface semimetal, in which two bands cross linearly on a 2D surface in the 3D Brillouin zone. This has been demonstrated in a family of graphene network materials, which all share a pair of almost flat Weyl surfaces at the Fermi level.\cite{Zhong2016} Note that a Weyl surface is distinct from the conventional Fermi surface, in that the surface intrinsically involves two bands and the low-energy quasiparticles are described by Weyl spinors. The stability of the Weyl surface will also need certain symmetry protections.\cite{Zhong2016}
\\
\\
\begin{figurehere}
\centerline{
\includegraphics[width=7.5cm]{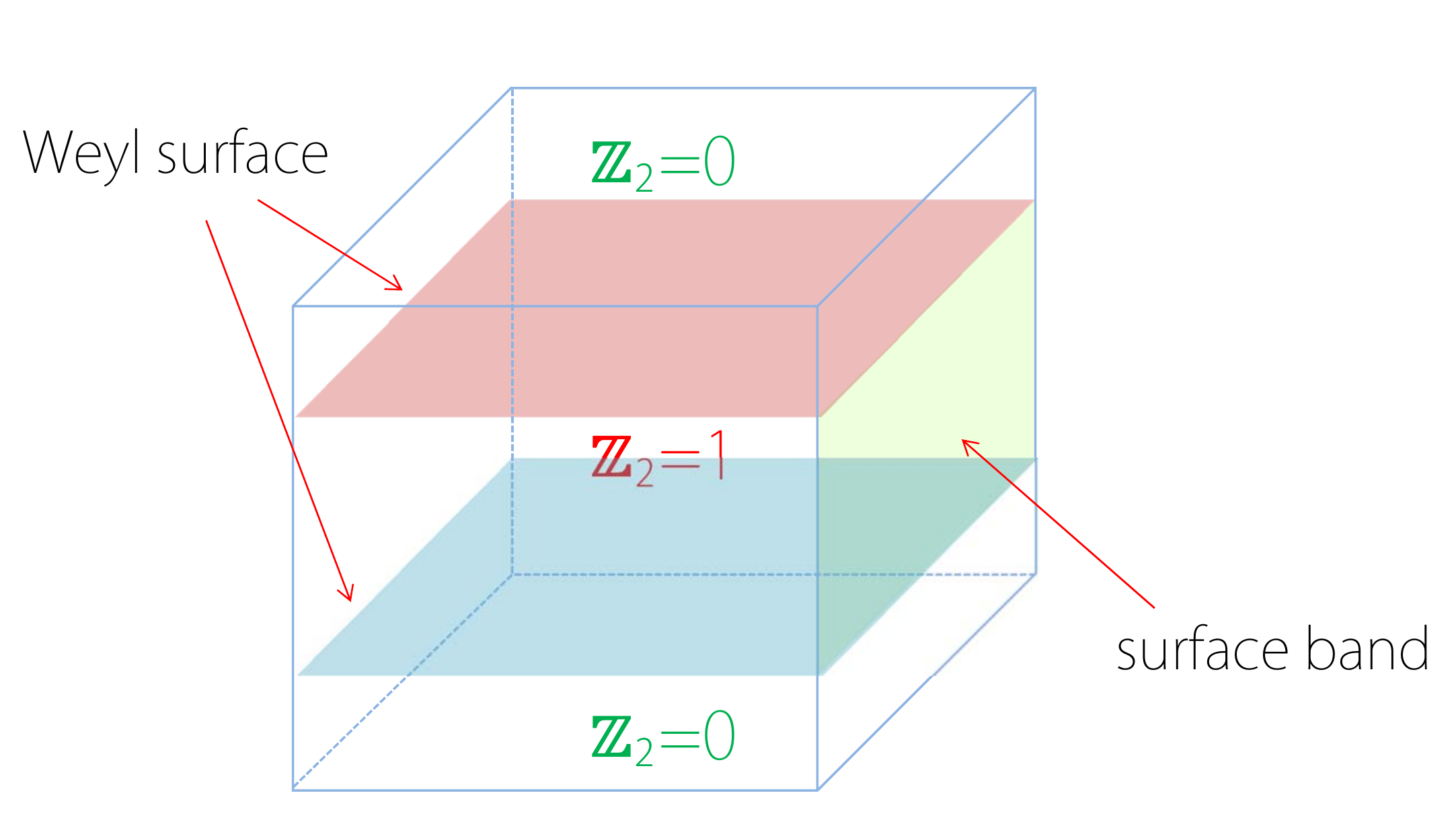}
}
\caption[]{Schematic figure of the Weyl surfaces in Ref.~\refcite{Zhong2016}. Each surface is formed by linear crossing of two bands (dispersion is linear
in the direction normal to the surface). The surfaces are stable as domain walls separating regions with inverted bandgaps. One the side surfaces, between the projections of the two Weyl surfaces, there exist almost flat surface bands close to the Fermi energy.}
\label{fig8}
\end{figurehere}
\begin{figurehere}
\centerline{
\includegraphics[width=7cm]{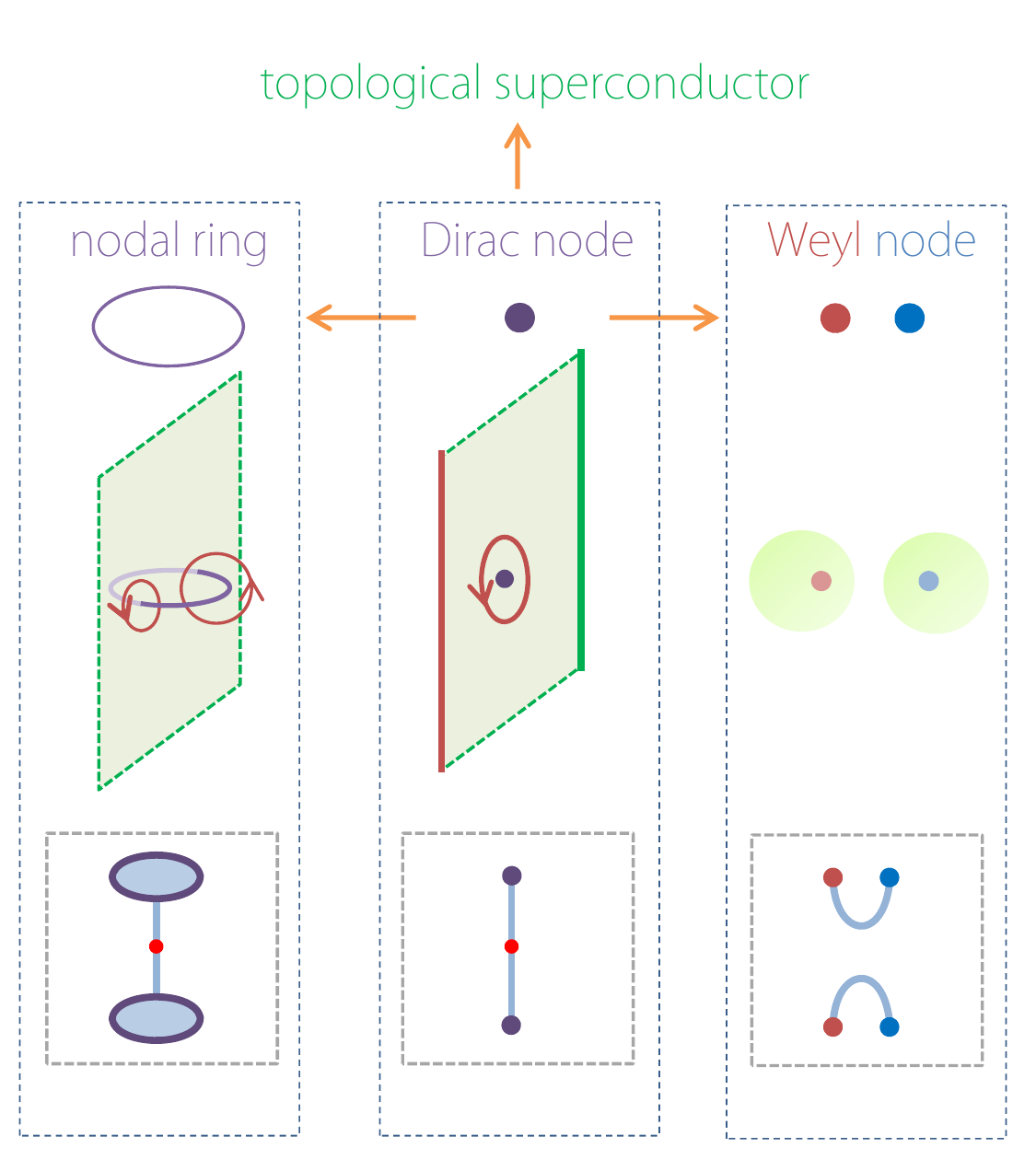}
}
\caption[]{3D Dirac superconductor in Ref.~\refcite{Yang2014}. Under breaking of inversion-gauge symmetry or time reversal symmetry, a Dirac node evolves into a nodal ring or two Weyl nodes respectively. The lower two panels in each dashed box show the topological protection of each zero energy object and the shape of surface zero modes. A Dirac superconductor may also be fully gapped to become a topological superconductor when mirror symmetry is broken. Figure adapted with permission from Ref.~\refcite{Yang2014}. Copyrighted by the American Physical Society.}
\label{fig9}
\end{figurehere}

Finally, we mention that the Dirac and Weyl concepts have also been generalized to superconductors. In fact, the nodal superconducting phases, topologically similar to the semimetal phases, are commonly seen in unconventional superconductors. Besides, for superconductors, there is an additional particle-hole symmetry for low-energy quasiparticle excitations, which may offer help to stabilize the nodal points. Weyl point like excitations have been proposed in the $^3$He-A phase,\cite{Volovik2009} in superlattice structures,\cite{Meng2012,Das2013} and in several bulk materials.\cite{Sau2012} The concept of Dirac superconductor was proposed in Ref.~\refcite{Yang2014}. Similar to Dirac points in semimetals, the superconducting Dirac node also requires protection from lattice symmetries. By controlling the symmetry breaking, one can transform the Dirac node into various topological objects, including nodal rings, Weyl nodes, and gapped topological superconductors, along with different surface zero modes (see Fig.~\ref{fig9}).\cite{Yang2014} A systematic analysis of the various nodal points in 3D chiral superconductors can be found in Ref.~\refcite{Kozii2016}.

\section{Some (Pseudo-)Spintronics Applications}

There could be multiple spin degrees of freedom in Dirac and Weyl materials. As mentioned, depending on the specific system, the Pauli matrices appearing in the Weyl Hamiltonian may or may not correspond to the real spin. In graphene, it is the sublattice orbital degree of freedom, and the real spin is largely uncoupled to the electron's motion. Whereas for the topological insulator surface, it is the real spin, making a strong coupling between real spin and orbital motion. The two scenarios represent two extreme cases, and each can find useful applications in spintronics, as reviewed by Pesin and MacDonald.\cite{Pesin2012} Here, instead, we shall be more focused on the manipulations with valley pseudospin. Sec. 3.1 and Sec. 3.2 will be discussing works on 2D and 3D materials, respectively.

\subsection{2D materials}

A prerequisite for spintronics applications is the ability to create spin polarization for carriers. Likewise, in order to utilize valley pseudospin, we need to have methods to generate valley polarization. At first glance, valleys are just different regions in the $k$-space, it seems unlikely that they could respond differently to external perturbations. Fortunately, for Dirac/Weyl materials, each valley has a definite chirality. There exist geometric electronic properties that are tied with the chirality and can be used to couple the valley pseudospin with external control fields.

The scheme was first discussed for graphene with inversion symmetry breaking.\cite{Xiao2007} As we have mentioned in Sec. 2.4, breaking of either $\mathcal{P}$ or $\mathcal{T}$ is necessary for a non-vanishing Berry curvature field. By still preserving $\mathcal{T}$, we could have the nice property that
$\bm \Omega(-\bm k)=-\bm \Omega(\bm k)$, such that the two valleys $K$ and $K'$, which are connected by $\mathcal{T}$, will acquire opposite Berry curvature fields. This makes the Berry curvature a valley-contrasting geometric property that can be used to differentiate the two valleys.
Another important valley-contrasting geometric property is the orbital magnetic moment:
\begin{equation}
\bm m_n(\bm k)=-i\frac{e}{2\hbar}\langle \nabla_{\bm k}u|\times [H(\bm k)-\varepsilon_n(\bm k)]|\nabla_{\bm k}u\rangle,
\end{equation}
where $\varepsilon_n$ is the band energy. This moment is entirely due to the orbital motion of Bloch electrons and has nothing to do with the spin magnetic moment (in fact, the spin moment may be understood as an effective $\bm m$ from Dirac equation\cite{Chuu2010}). For a two-band model, one can show that $\bm m$ and $\bm \Omega$ are connected through the relation
\begin{equation}
\bm m_n (\bm k)=\frac{e}{2\hbar}[\varepsilon_n(\bm k)-\varepsilon_{n'}(\bm k)]\bm \Omega_n(\bm k),
\end{equation}
where $n,n'=\pm$, with $+$ and $-$ refer to the conduction (higher) and valence (lower) bands respectively, and $n\neq n'$. Consequently, $\bm m$ also becomes valley contrasting like $\bm \Omega$ for Dirac valleys. For the Dirac model with inversion symmetry breaking,
\begin{equation}
H(\bm k)=v(\tau_z k_x\sigma_x+k_y\sigma_y)+\Delta\sigma_z,
\end{equation}
we have (hereafter we set $\hbar=1$)
\begin{equation}
\bm\Omega_n (\bm k)=-\tau_z n\frac{v^2\Delta}{2(\Delta^2+v^2 k^2)^{3/2}}\hat z,
\end{equation}
\begin{equation}
\bm m_n(\bm k)=-\tau_z\frac{v^2 m_e \Delta}{(\Delta^2+v^2 k^2)}\mu_B\hat z,
\end{equation}
where $m_e$ is the electron mass, and $\mu_B=e\hbar/2m_e$ is the Bohr magneton. These expressions explicitly show that these geometric properties are tied with the chirality and the valley pseudospin $\tau_z$. Both quantities are peaked at the valley center $k=0$ and quickly decay as $k$ increases. $\bm \Omega$ vanishes when $\Delta=0$, i.e. when the symmetry is unbroken, as consistent with our previous assertion. (Note that for this case, the band crossing point at $k=0$ is a singular point for $\bm \Omega$, which is the source for the $\pi$ Berry phase in Eq.(\ref{BPhase}). This degenerate point requires a non-Abelian treatment.\cite{Culcer2005,Xiao2010}) One also notes that in 2D, both $\bm \Omega$ and $\bm m$ only have a single component perpendicular to the plane.

How would these geometric properties affect the electron dynamics? This can be best appreciated in the semiclassical equations of motion for an electron wave-packet in a single band:\cite{Sundaram1999}
\begin{equation}\label{rdot}
\dot{\bm r}=\frac{\partial \mathcal{E}}{\partial \bm k}-\dot{\bm k}\times\bm \Omega,
\end{equation}
\begin{equation}
\dot{\bm k}=-e\bm E-e\dot{\bm r}\times \bm B,
\end{equation}
where $\mathcal{E}=\varepsilon_n(\bm k)-\bm m(\bm k)\cdot\bm B$. Therefore, the Berry curvature $\bm \Omega$, like a magnetic field in $k$-space, generates an anomalous velocity term, in correspondence with the Lorentz force from $B$-field; meanwhile the orbital magnetic moment $\bm m$ couples with the $B$-field as like a Zeeman term, producing a shift of the band energy. This set of equations of motion are accurate to the first order in external fields, and a generalization to second order has been done in Ref.~\refcite{Gao2014}.

Due to the anomalous velocity term, an in-plane $E$-field will drive a transverse carrier flow. Because $\bm \Omega$ takes opposite signs for the two valleys, with finite doping, carriers at the two valleys flow in opposite directions, giving rise to a pure valley current. This is the valley Hall effect predicted by Xiao, Yao, and Niu.\cite{Xiao2007} These chirality-dependent quantities also directly enter into the optical matrix element for coupling with circularly polarized light (can be thought of a chiral state of light). Yao \emph{et al.} proposed that a circularly polarized light can be used to selectively excite one of the valleys.\cite{Yao2008} From the valley-dependent Zeeman-like coupling $-\bm m\cdot \bm B$, $B$-field produces different energy shift for the two valleys. Based on this effect, Cai \emph{et al.} proposed the control of valley polarization by a magnetic field.\cite{Cai2013} Thus, with the help of these valley-contrasting geometric properties, one could generate valley polarization through electric, optic, and magnetic means.

These mechanisms were originally proposed in the graphene context, and experimentally realized first in the TMD materials.\cite{Mak2012,Zeng2012,Cao2012,Xu2014d,Mak2014} They are generic to valleys with non-vanishing Berry curvatures hence are expected to be useful also for other types of Dirac/Weyl materials.

\begin{figure*}
\begin{center}
\includegraphics[width=12cm]{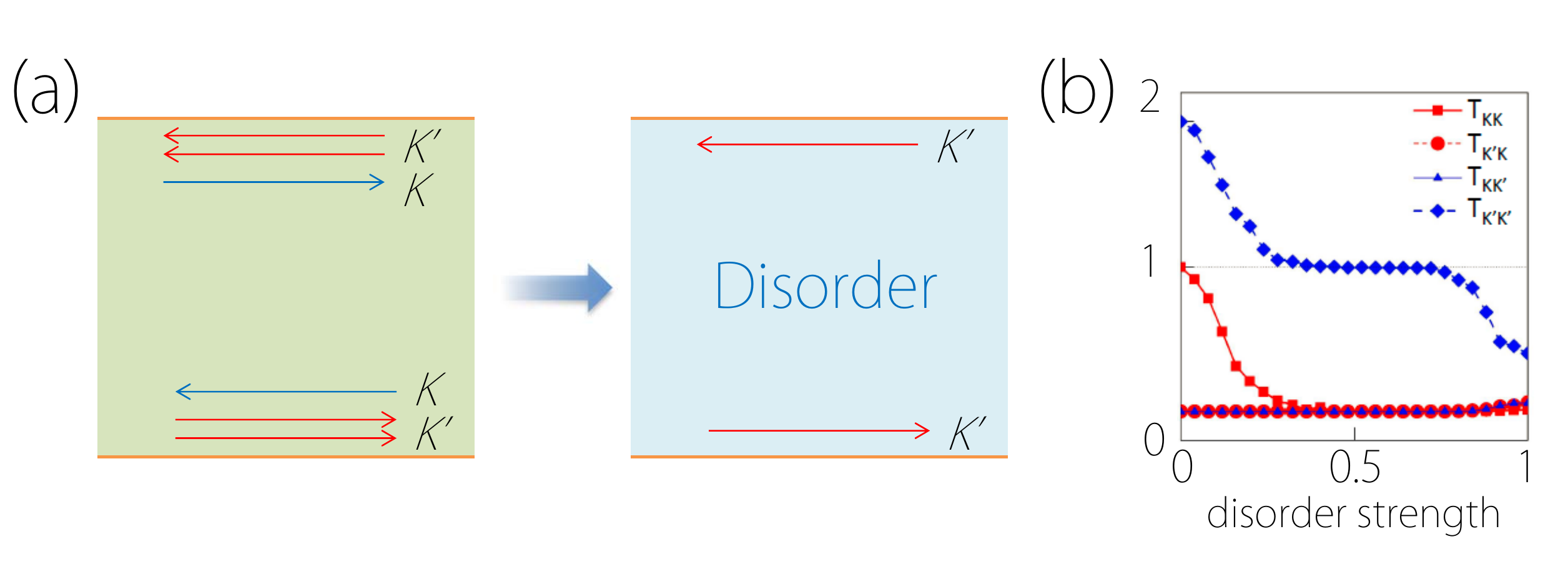}
\end{center}
\caption[]{(a) When numbers of edge channels in the two valleys are unbalanced, by adding short-range scatterers, pairs of counter-propagating channels are annihilated, making the system an effective medium with completely valley-polarized channels. (b) Numerical simulation results for the valley-resolved transmission probability as a function of disorder strength for a two-terminal setup, showing a plateau for transmission in $K'$ valley while other processes are suppressed. Figure adapted with permission from Ref.~\refcite{Pan2015}. Copyrighted by the American Physical Society.}
\label{fig10}
\end{figure*}

After creating the valley pseudospin polarization, we still need ways to process the valley information. Analogous to spin filters in spintronics, it is desired to have a valley filter device which can selectively filter out one pseudospin species. Generally, a valley filter should have a channel region with the property that it only allows one kind of valley to pass through.
The concept was first proposed by Rycerz \emph{et al.} in graphene.\cite{Rycerz2007} They found that a graphene nanoconstriction with zigzag edge can be controlled to pass only one valley and strongly reflect the other valley. Following this work, various other schemes are proposed by using irradiation,\cite{Abergel2009} line defects,\cite{Gunlycke2011} or strain engineering.\cite{Jiang2013,Fujita2010}
\begin{figure*}
\begin{center}
\includegraphics[width=11cm]{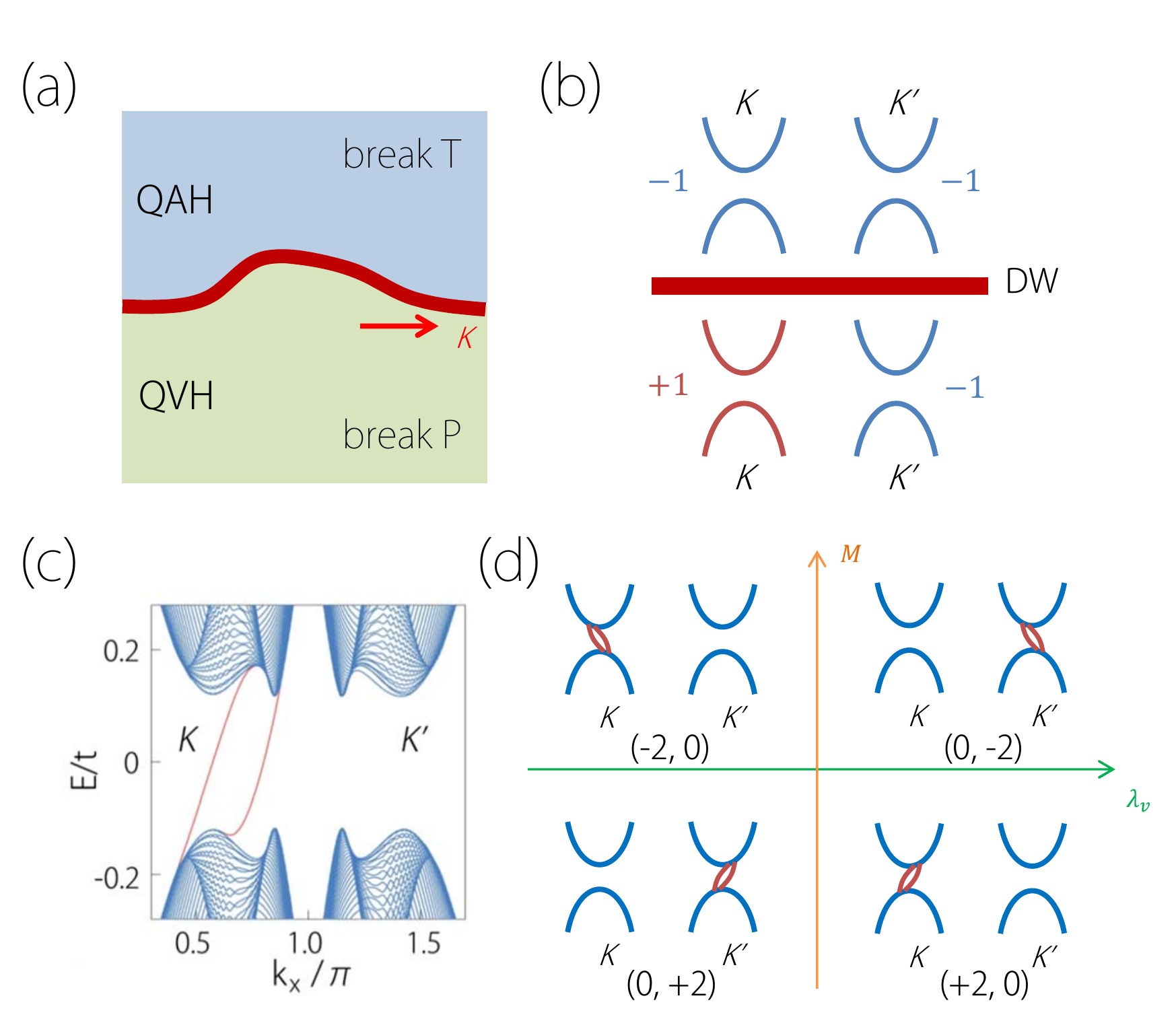}
\end{center}
\caption[]{(a) Perfect valley filter formed at the domain wall between the quantum anomalous Hall and the quantum valley Hall domains. (b) and (c) are results from a single layer graphene model. The number of valley channels in the gap are determined by the difference of valley topological charges across the domain wall. In the current case, there are two channels in the $K$ valley and no channel in the $K'$ valley. (d) Valley index and propagating direction (chirality) of the channels can be controlled by tunable model parameters. Figure adapted with permission from Ref.~\refcite{Pan2015a}. Copyrighted by the American Physical Society.}
\label{fig11}
\end{figure*}

One recent proposal is based on the so-called valley polarized quantum anomalous Hall (QAH) phase.\cite{Pan2015,Zhang2011} In this phase, there exist topologically protected 1D channels on the sample edge, more importantly, these channels are valley polarized with unbalanced numbers.\cite{Pan2014} For example, in the case shown in Fig.~\ref{fig10}(a), there is one channel for the $K$ valley and two channels for the $K'$ valley on the upper edge. Due to their chiral nature, only the $K'$ valley carriers can pass from right to left on the upper edge. Nevertheless, the $K$ valley carriers can still pass through the system, by going around the lower edge. It was found that the valley filtering can be achieved by introducing short-range scatterers.\cite{Pan2015} Note that for most valleytronics applications, sharp scatterers  should usually be avoided because they scatter carriers to different valleys and shorten the valley lifetime. However, in the current case, from numerical calculations, it was demonstrated that the scattering helps to annihilate a pair of edge channels with opposite chirality, and effectively leave the system with a single chiral channel belonging to the $K'$ valley (Fig.~\ref{fig10}). Then the system can be used as a filter, only allowing $K'$ carriers to pass through.

In most valley filter designs, there are propagating modes from both valleys in the channel region. Carriers can be scattered between these modes due to lattice-scale defects, which are difficult to be completely eliminated. In the above proposal using edge channels, the control of scatterers would also be a challenge. In a recent work, the concept of a ``perfect valley filter" was proposed.\cite{Pan2015a} In the channel region of such filter, there exist topologically-protected channels residing in only one valley, and propagating in only one direction (e.g. see Fig.~\ref{fig11}(c)). In such a filter, there is in principle no intervalley scattering due to the absence of propagating modes in the other valley (unless for inelastic scattering involving large energy transfer outside the bandgap). And the carriers, once entering the filter, cannot be backscattered because all the propagating modes are unidirectional. Furthermore, as long as the valley character can be differentiated at the entrance and the exit of the filter (i.e. the valley projections onto the 1D channel are fairly separated), the middle part of the channel can take arbitrary shapes, which will not affect its performance.

It was proposed that such perfect filter can be realized at the boundary (domain wall) between two topological domains, one QVH domain with broken $\mathcal{P}$ and one QAH domain with broken $\mathcal{T}$ (Fig.~\ref{fig11}).\cite{Pan2015a} It was noted that for each valley, the number of chiral channels $\nu$ is determined by the difference of so-called valley topological charge $N$ across the domain wall,\cite{Martin2008,Yao2009}
\begin{equation}
\nu=N(\mathrm{domain\;1})-N(\mathrm{domain\;2}),
\end{equation}
where $N$ can be expressed as an integral of Berry curvature around the valley. Hence the desired configuration can be achieved by making $\nu\neq 0$ for one valley and $\nu=0$ for the other valley. Specific realizations in single layer and bilayer graphene are suggested.\cite{Pan2015a}

\subsection{3D materials}

Many works on 3D Weyl semimetals are focused on the chiral anomaly, an effect originally proposed in high energy physics. It states that under parallel electric and magnetic fields, electrons can be pumped from Weyl points with one chirality to points with the opposite chirality.\cite{Nielsen1983} Experimentally, this can be detected as a negative magnetoresistance signal.\cite{Huang2015a,Zhang2016,Arnold2016} If the multiple Weyl points can be viewed as a pseudospin degree of freedom, the chiral anomaly effect would provide a way to generate a pseudospin polarization.

More directly, since the 3D Weyl points are like source charges for the Berry curvature field, like in the 2D case, the electron dynamics will depend on the chirality of the point due to the Berry curvature as in Eq.(\ref{rdot}). Under a driving force, the electron orbit will be deflected.

The effect is most striking when considering the scattering of a Weyl electron at a potential step.\cite{Jiang2015a,Yang2015b} Due to the anomalous velocity term, the reflected (transmitted) electrons acquire a shift $\delta y^R$ ($\delta y^T$) as in Fig.~\ref{fig12}(a), that is normal to the incident plane. A simple argument of this effect can be provided in terms of angular momentum conservation.\cite{Yang2015b} As stated in Eq.(\ref{AM}), a Weyl fermion wave-packet has a total angular momentum,
\begin{equation}
\bm J=\bm r\times\bm k+\frac{\tau_z}{2}\bm n,
\end{equation}
where $\tau_z$ is the chirality and $\bm n$, the (pseudo-)spin direction, is tied with the particle's momentum. Assuming rotational symmetry in the interface $x$-$y$ plane, $J_z$ is a conserved quantity. For both transmission and reflection, the particle's momentum is changed (see Fig.~\ref{fig12}(b)), so is $\bm n$. To maintain a conserved $J_z$, there has to be a shift in the orbital part to compensate the change in the spin angular momentum. And due to the $\tau_z$ dependence, the shift will be opposite for the two chirality, so the effect is termed as the ``chirality Hall effect".\cite{Yang2015b} The result obtained using this argument agrees with that from the semiclassical equations of motion and from the scattering approach. It can be viewed as an electronic analog of the Imbert-Fedorov shift in optics.\cite{Fedorov1955,Imbert1972,Onoda2004} Like valley Hall effect in 2D, it may be used to design devices for separating carriers according to their chirality, but one should note that the effect is an intrinsically 3D effect, because the shift is in the third direction perpendicular to the plane of motion.
\begin{figurehere}
\centerline{
\includegraphics[width=8cm]{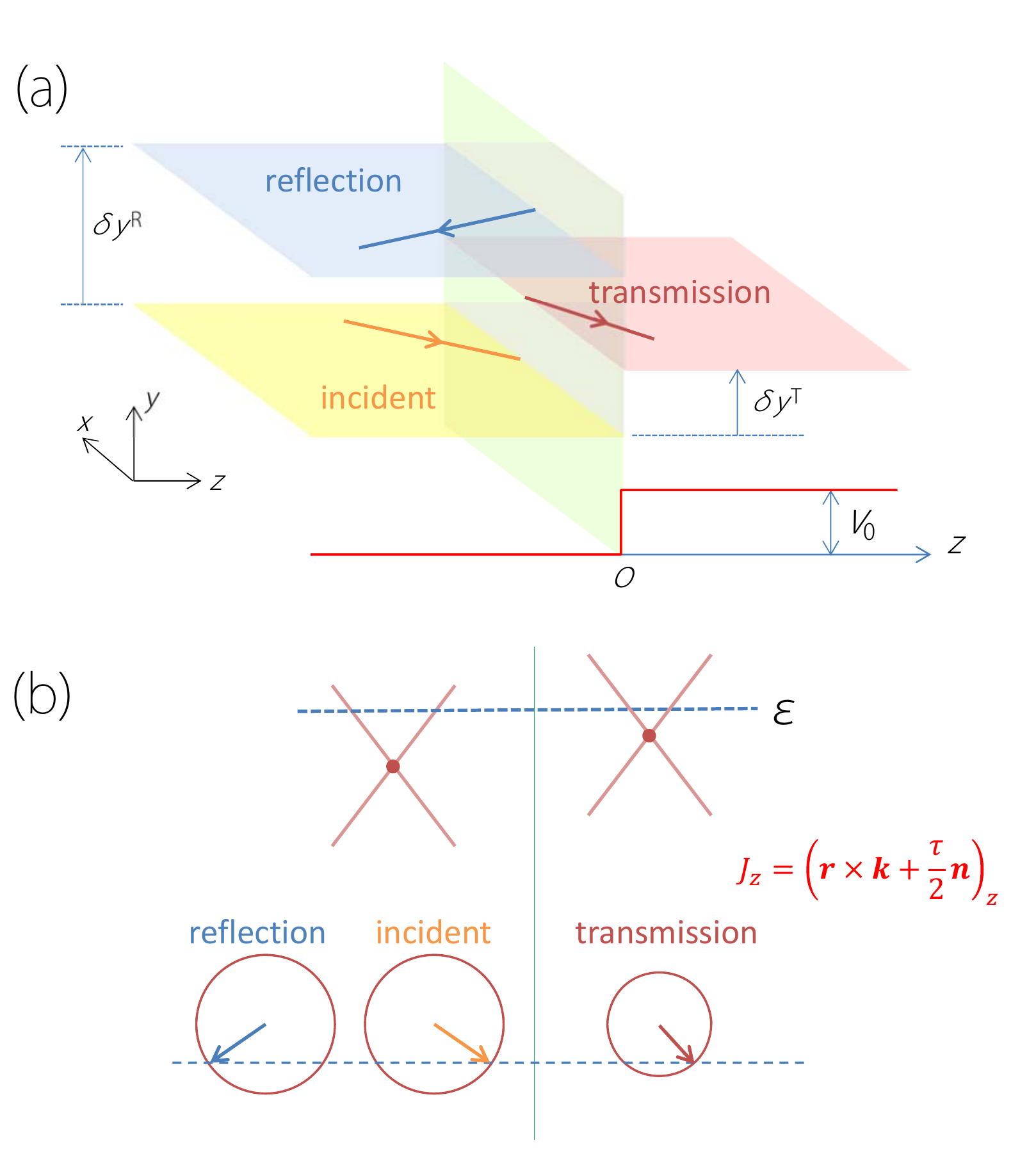}
}
\caption[]{(a) A 3D Weyl fermion incident on a potential step acquires a shift normal to the incident plane for both reflection and transmission. (b) The Weyl fermion's wave vector changes for transmission and reflection, which alters its spin angular momentum. With conserved total angular momentum, the orbital part must compensate for the change, leading to the transverse shift. Figure adapted with permission from Ref.~{\refcite{Yang2015b}}. Copyrighted by the American Physical Society. }
\label{fig12}
\end{figurehere}

For device applications, it is in fact more convenient to work with nanostructures instead of bulk materials. With confinement along one direction, we can have a quasi-2D system. This scheme is particularly useful for the Dirac semimetal materials Na$_3$Bi and Cd$_3$As$_2$ due to their simple Dirac point configuration.\cite{Hellerstedt2016} Consider the confinement along $z$-direction. The electron's motion in $z$-direction is quantized into quantum well subbands, with an effective wave vector of $k_z\approx n\pi/L$ with $n$ a positive integer and $L$ is the confined length in $z$.\cite{Xiao2015} As we discussed after Eq.(\ref{DSMH}), the 2D slice with a specific $k_z$ can have a nontrivial $\mathbb{Z}_2$ invariant only when $k_z<k_D$. Hence depending on its effective $k_z$, a subband can be topologically trivial or nontrivial. If an odd number of subbands are nontrivial, the whole system will be in a QSH insulator phase; otherwise the system is trivial. This leads to an interesting oscillation of the topological invariant as the thickness $L$ increases.\cite{Wang2013b,Xiao2015}
\\
\\
\begin{figurehere}
\centerline{
\includegraphics[width=8cm]{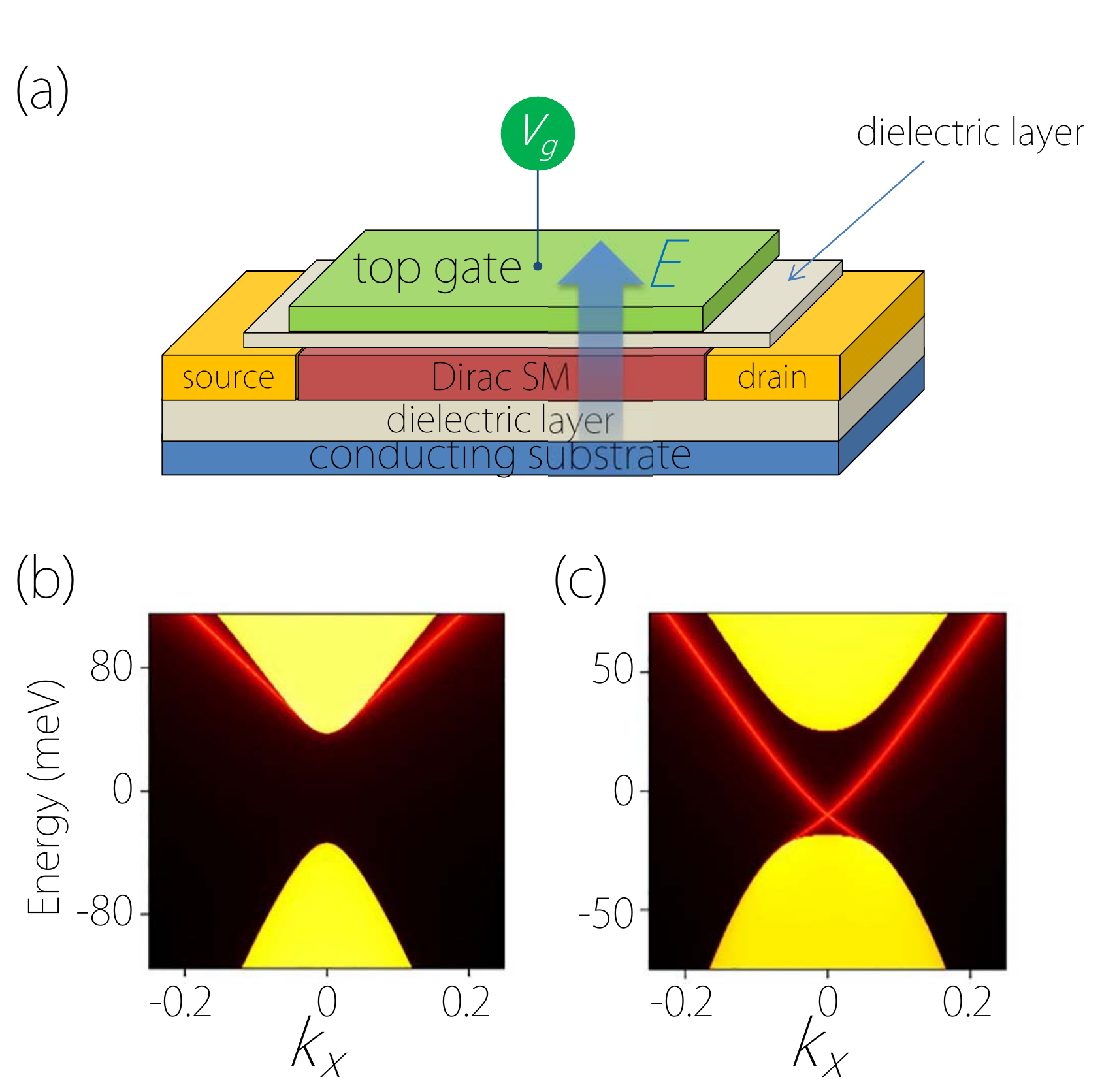}
}
\caption[]{(a) Schematic of a topological field effect transistor using Dirac semimetal as channel material. The conduction is through the topological edge channels, controlled through the field-induced phase transition. (b) and (c) are the energy spectrum calculated for Na$_3$Bi thin films with gate $E$-field off and on, respectively. The absence and presence of edge channels can be observed. Figure (b) and (c) adapted with permission from Ref.~{\refcite{Pan2015b}}.}
\label{fig13}
\end{figurehere}

With nanoscale confinement, a sizable $E$-field effect can be generated by the standard gating technique. For example, with transverse gating, one can control the Rashba-type spin splitting of the topological edge channels.\cite{Xiao2015} More interestingly, a vertical $E$-field can be used to tune the gap of the subband for a Dirac semimetal thin film (Fig.~\ref{fig13}(a)).\cite{Pan2015b} It can induce a gap inversion and change the topological $\mathbb{Z}_2$ character of the subband. As a result, the system as a whole will undergo a field-induced topological phase transition between a trivial insulator and a QSH insulator. Importantly, this controls the presence/absence of the spin-helical edge channels to conduct charge and spin currents (see Fig.~\ref{fig13}(b,c)). Therefore, the Dirac semimetal thin film can be used to fabricate a topological field effect transistor.\cite{Pan2015b}

Besides electric field, a recent work by Guan \emph{et al.} shows that the topological phase transition between the QSH and trivial insulator phases can also be efficiently controlled by a small lattice strain.\cite{Guan2016} This makes it possible to realize a novel ``piezo-topological transistor" device.
Furthermore, with ferromagnetic source and drain contacts, the system can also be turned into an efficient spin polarization rotator or spin modulator.\cite{Xiao2016} Because of the chiral character of the edge channels, backscattering can be minimized. It is believed that such devices can enjoy the advantages of fast speed, low power consumption, and low heat dissipation.

\section{Summary}

In this paper, we have introduced the basic concepts in the field of Dirac and Weyl materials, and discussed some recent works on the generalization of the concepts towards different directions and on the attempt to use the unique properties of these materials for applications. Due to the limited length, many important and exciting progress cannot be covered. We hope that the paper at least served its humble purpose, i.e. to demonstrate that crystalline solids offer a fertile ground for exploring various types of emergent fermions, some having their high energy counterparts and more even going beyond; and these new materials possess a wealth of intriguing properties distinct from conventional metals or semiconductors, which are expected to attract more and more interest for fundamental research as well as technological applications.

\section{Acknowledgement}

The author thanks Ying Liu, Shan-Shan Wang, Zhi-Ming Yu, Fan Zhang, Wenye Duan, J. Xiao, and D.L. Deng for valuable discussions, and thanks Shan Guan for help with the figures. The work was supported by
Singapore MOE Academic Research Fund Tier 1 (SUTD-T1-2015004) and by SUTD-SRG-EPD2013062.

\bibliographystyle{ws-spin}
\bibliography{SPIN-Rev}

\end{multicols}
\end{document}